\documentclass[a4paper,9pt]{article}

\usepackage{graphicx}
\usepackage{mathrsfs}
\usepackage{amssymb}
\usepackage{amsmath}
\usepackage{bm,latexsym}
\newcommand{\be}{\begin{equation}}
\newcommand{\ee}{\end{equation}}
\newcommand{\bea}{\begin{eqnarray}}
\newcommand{\eea}{\end{eqnarray}}
\newcommand{\ena}{\end{eqnarray}}

\usepackage{graphicx,amssymb,amsmath,bm,latexsym,color}

\newcommand{\vs}[1]{\vspace{#1 mm}}

\newcommand{\hA}{{\hat A}}
\newcommand{\m}{{\hat m}}
\newcommand{\hr}{{\hat r}}

\begin{document}

\begin{titlepage}

\vspace{.5in}
\vskip .5in

\begin{center}

{\Large\bf Creation of a black hole pair with a domain wall}
\vskip .5in

{\large
Bogeun Gwak $^{a}$ \footnote{\it email:rasenis@sogang.ac.kr},
Bum-Hoon Lee $^{a,b}$ \footnote{\it email:bhl@sogang.ac.kr},\\
{Wonwoo Lee$^{b}$ \footnote{\it email:warrior@sogang.ac.kr},}
Masato Minamitsuji$^{c}$ \footnote{\it email:masato.minamitsuji@kwansei.ac.jp} }
\\
\vs{10}
{$^a${\em Department of Physics and BK21 Division, Sogang University,
Shinsu-dong 1, Mapo-gu, Seoul, 121-742, Republic of Korea} } \\
$^b${\em Center for Quantum Spacetime, Sogang University,
Shinsu-dong 1, Mapo-gu, Seoul, 121-742, Republic of Korea} \\
$^c${\em Department of Physics, School of Science and Technology,
Kwansei Gakuin University,
Sanda, Hyogo 669-1337, Japan}

\begin{abstract}
We study the creation of a black hole (BH) pair separated by a domain wall, in the presence of a cosmological constant. We construct the solution representing a BH pair with a domain wall and compute the Euclidean action to evaluate the probability of the pair creation in the background with a preexisting domain wall. The BHs can be either neutral or magnetically charged ones. We compare the results of the charged case with those of the neutral case with the same cosmological constant. We find that the production rate of a charged BH pair is always suppressed in comparison with that of the neutral one in both four and five dimensions, irrespective of the sign of the cosmological constant. The Euclidean action is equal to the minus of the entropy. Since the horizon area of a BH is decreased as the magnitude of its charge is increased in general, the decreasing creation rate can be understood in terms of the increasing charge. We obtain the explicit confirmation on the relation between
the pair creation rate of the charged BHs and the area of horizons
in both the four- and five-dimensional cases in the presence of a cosmological constant. The singularity of the domain wall universe with charged BHs, as distinct from that with neutral BHs, can be avoided.


\end{abstract}

\end{center}

\end{titlepage}

\newpage
\setcounter{page}{1}
\section{Introduction}

Black holes (BHs) are the fundamental objects in studying gravity, astrophysics and even high energy physics. The astrophysical BHs are
the most promising sources of the gravitational waves and important for testing the general relativity in strong gravity regime.
In cosmology, the primordial BHs will be probes
for cosmic history and structure formations \cite{carr}.
Furthermore, recent developments in higher-dimensional theories
have brought our interests in finding new higher-dimensional BH solutions as well as other kinds of black objects, and
investigating the possibility of their productions in the high energy colliders as LHC \cite{lhcgt, scp, cmsc}.

The vacuum in the presence of a strong field can decay through the so-called Schwinger mechanism \cite{schwinger}, creating a pair of particles in an external electric field. Such particle production is one of the decaying processes of the background field in the given vacuum state (for recent works, see e.g., \cite{kly02, hag}). As the gravitational analogy of the mechanism, it was proposed that the pair creation of BHs, which represents a nonperturbative topological fluctuation of the gravitational field, could be possible in the background magnetic field \cite{gib01}. The energy needed for the creation is provided by the background field. This was confirmed in \cite{giga01} by constructing an appropriate Ernst instanton solution which describes a pair creation of BHs. The BH pair creation involving a cosmic string has been studied \cite{hr06, emp01, odl01, od00}. In this case, the energy to create the BH pair comes from the string tension. The pair creation has been also extensively investigated in the early universe \cite{bh001, bh003, bh002, od02}, where the energy comes from the cosmological constant, and in other frameworks \cite{dggh1, bm00, wu01, gkp0, cm41}. In the case with a domain wall, the wall's gravitational repulsive energy \cite{is01} can give rise to a pair creation of BHs. Such a process was first studied in \cite{caldwell} using a cut-and-paste method. The topological BHs in the presence of the domain wall \cite{mann000}, and the creation of an inflationary braneworld with a pair of black cigars \cite{gasa} have been studied.

There are two kinds of approach to the creation of a BH pair. One is the bounce approach. In this approach, one needs to find an instanton solution of the Euclidean field equations, which interpolates the initial state without BHs and the final state with BHs. The other is the quantum cosmological approach. In this approach, one can employ the no-boundary proposal \cite{jhsh} which represents the probability of the final state from nothing and that of the initial state from nothing. The creation rate from the initial to the final state can be obtained from these two probabilities. This method does not need the interpolating instanton solution. In the semiclassical approximation, the formula of the production rate without a prefactor in the bounce approach is equivalent to the formula in the quantum cosmological approach \cite{bcha}.

In this paper, we revisit the issue of the creation of a BH pair separated by a domain wall with an arbitrary cosmological constant. One of our motivations is the expectation that the creation of the domain wall with a charged BH pair in the five-dimensional spacetime may provide a model of the braneworld universe. In terms of string theory or supergravity, the theory generically contains higher-rank antisymmetric tensor fields and BHs may be charged. Thus it is naturally expected that in any string-inspired model the domain wall universe is created together with the charged BHs. In addition, from the cosmological points of view, the possibility of the domain wall universe with charged BHs is worth being considered since the initially contracting universe can experience a bounce
and the singularity can be avoided. From these points of view, it is very significant to evaluate the nucleation rate
of the charged BHs. Thus, we could get insights on the quantum cosmological origin of braneworlds. We will construct the solution representing the charged and the neutral BH pair with a preexisting domain wall in the four- and five-dimensional spacetimes. We will compute their production rates, and then compare the actions of the charged case with those of the neutral case with a same cosmological constant rather than obtaining the interpolating instanton solutions.

The spacetime of the preexisting wall can be provided either by a cut-and-paste method or by the instanton solution mediating tunneling between the degenerate vacua in curved space \cite{hw, Lee:2008hz, lllo, lllo2}. The braneworld-like object can be obtained as the interpolating instanton solution by applying the mechanism in \cite{lllo, lllo2}. From this point of view, our present work can be the basic framework to make the spacetime, where we can describe the dynamics of a domain wall universe in the charged BH spacetime not only in the  Einstein-Maxwell theory in Refs.~\cite{mp01, km01} but also in the more general $U(1)$ gauge theories in Refs.~\cite{llm01, mh00, dm01}.

The organization of this paper is as follows: In Sec. 2, we review the magnetically charged BHs for general spacetime dimensions in the presence of a cosmological constant. In Sec. 3, we construct the solution for a charged BH pair separated with $Z_2$ symmetry by a domain wall in the four- and five-dimensional spacetimes. The case of a neutral BH pair including non-$Z_2$ symmetry was studied in \cite{Lee:2008hz} using the Bousso-Hawking normalization \cite{bh003, bh002} in the bulk, where ambiguity of the time periodicity on the wall remains. In Sec. 4, we evaluate the Euclidean action and the resultant pair creation rate for the above system using the time periodicity by the Hawking temperature of BHs \cite{hh00} in four- and five- dimensional spacetimes. We will show that the creation rate of a BH pair with the domain wall is proportional to the area of the horizons of created BHs in four- and five-dimensional spacetimes. We will obtain the first explicit confirmation on the relations between the pair creation rate and the area of horizons. And the application to the braneworld cosmology is discussed. In Sec. 5, we summarize our results and discuss the possible generalizations of our studies. It is natural to expect that the same properties are hold also in more than six-dimensional spacetime which will be discussed.

\section{Magnetic BHs }

We consider a system composed of Einstein gravity with either positive or negative cosmological constant $\Lambda$,
coupled to the antisymmetric tensor field strength of rank $(n-1)$ $F_{(n-1)}$
\bea
S&=&\int_M d^{n+1} x\sqrt{-g}
\left[
\frac{1}{2\kappa^2}\Big(R-2\Lambda\Big)
-\frac{1}{2 (n-1)! }
F_{\alpha_1 \cdots \alpha_{n-1}}F^{\alpha_1 \cdots \alpha_{n-1}} \right] \nonumber \\
&+& \oint_{\partial M} d^n x \sqrt{-q} \left[
\frac{K-K_o}{\kappa^2} + \frac{1}{(n-2)!} F^{\alpha_1 \cdots \alpha_{n-1}}n_{[\alpha_1} A_{\alpha_2 \cdots \alpha_{n-1}]}\right] ,
\eea
where $g\equiv \mathrm{det} g_{\mu\nu}$, $q\equiv \mathrm{det} q_{ab}$, $K$ and $K_{o}$ are traces of the extrinsic curvatures of ${\partial M}$ in the metric $g_{\mu\nu}$ and $\eta_{\mu\nu}$, respectively, $F_{\alpha_1 \cdots \alpha_{n-1}}= (n-1) \nabla_{[\alpha_1} A_{\alpha_2 \cdots \alpha_{n-1}]}$, and $n_{\alpha_1}$ is a unit normal vector which points outward to the boundary. The Greek indices run the $(n + 1)$-dimensional spacetime, while the Roman indices run the $n$-dimensional spacetime. The third term on the right-hand side corresponds to the boundary term for Einstein gravity \cite{ygh}.  The fourth term corresponds to the boundary term for the bulk form field. The term is not needed in the case of a magnetic field. In the electric case, there is the subtlety that the field must be purely imaginary on the Euclidean section and we must keep the boundary term for the bulk part \cite{rhpb}. Note, however, the partition function in a definite charge sector for the electric cases with the boundary term is the same as that for the magnetic cases without the boundary term in the semiclassical approximation. Hereafter we will omit the boundary term for the form field, because it will vanish in this paper.

Varying the action with respect to the metric gives the Einstein equation
\bea
R_{\mu\nu} =\frac{2\Lambda}{n-1}g_{\mu\nu} +\frac{\kappa^2}{(n-1)!}
\Big[(n-1)\big(F_{[n-1]}^2\big)_{\mu\nu} -\frac{n-2}{n-1}g_{\mu\nu}\big(F_{[n-1]}^2\big) \Big],
\eea
where we have defined
$\big(F_{[n-1]}^2\big)_{\mu\nu}= F_{\mu}{}^{\alpha_{1}\cdots \alpha_{(n-2)}}F_{\nu\alpha_{1}\cdots \alpha_{(n-2)}}$
and $\big(F_{[n-1]}^2\big)$ is its trace.

A magnetically charged BH solution in the (anti-) de Sitter ((A)dS) spacetime is given by
\bea
\label{metrics}
&&ds^2=-f(r) dt^2+\frac{dr^2}{f(r)}+r^2 d\Omega_{(n-1)}^2,
\quad
f(r)=1-\frac{2 \Lambda r^2}{n(n-1)}
      -\frac{2m}{r^{n-2}}
      +\frac{q^2}{r^{2(n-2)}},
\nonumber\\
&&
F_{a_1a_2\cdots a_{n-1}}=
\frac{\sqrt{(n-1)!}q}{\kappa}
\sqrt{\gamma_{(n-1)}}
\epsilon_{a_1\cdots a_{n-1}},
\quad
\eea
where $\epsilon_{a_1\cdots a_{n-1}}=\pm 1$, $\gamma_{(n-1)}$ is the volume element of the $(n-1)$-sphere, and indices $\{a_i\}$ are those for it. The other components of the form field are zero. The existence and number of the horizons
crucially depend on the parameters. In this paper, we will focus on the cases of four- and five-dimensional BHs.

\subsection{The four-dimensional case}

In the case of $n=3$, the solution Eq.\ (\ref{metrics}) becomes the magnetic Reissner-Nordstroem-(A)dS BH in four dimensions
\bea
&&ds^2=-f(r) dt^2+\frac{dr^2}{f(r)}+r^2
\big(d\theta^2+\sin^2\theta d\phi^2\big),
\quad
f(r)=1-\frac{\Lambda r^2}{3}
      -\frac{2m}{r}
      +\frac{q^2}{r^{2}},
\nonumber\\
&&
F_{\theta\phi}=
\frac{\sqrt{2}q}{\kappa} \sin \theta.
\eea
\begin{figure}
   \begin{center}
    \includegraphics[scale=.80]{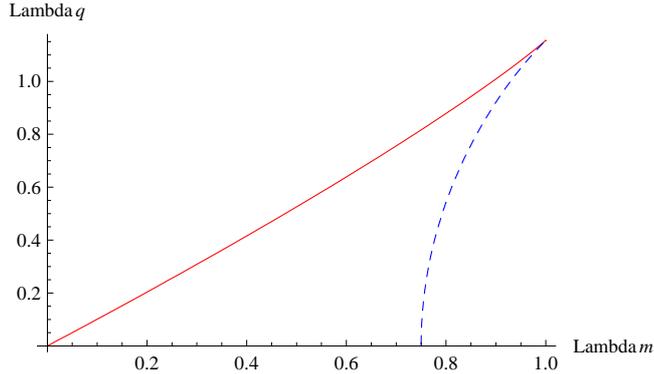}
        \caption{ The phase of a magnetically
charged BH with a positive cosmological constant
in five-dimensional spacetime is shown. The vertical and horizontal axes show $\Lambda q$ and $\Lambda m $, respectively. The solid (red) curve corresponds to the extremal solutions $r_+=r_-$ (Eq. (\ref{ext})), while the dashed (blue) curve corresponds to the
charged Nariai solutions $r_+=r_c$ (Eq. (\ref{cN})).
The intersecting point of these curves corresponds to
the ultracold solution $r_+=r_-=r_c$.}
   \end{center} \label{fig1}
\end{figure}
\begin{figure}
   \begin{center}
    \includegraphics[scale=.80]{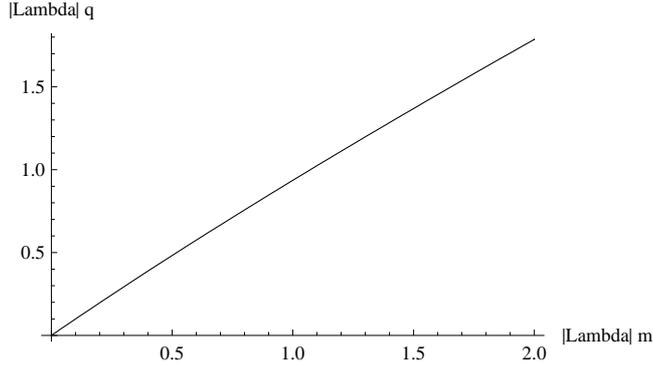}
        \caption{
The phase of a magnetically charged BH
with a negative cosmological constant
in five-dimensional spacetime
is shown.
The vertical and horizontal axes show $|\Lambda|q $
and $|\Lambda|m $.
The solid curve corresponds to the extremal solutions $r_+=r_-$.
For solutions corresponding to the side below this curve,
there are two horizons, and for those left to this curve
there is no horizon.}
   \end{center}\label{fig2}
\end{figure}

There are at most two BH horizons $r_{\pm}$ ($r_+\geq r_-$).
There is also a cosmological horizon outside the BH horizons $r_c\geq r_+$. The phase is quite similar to Fig.\ 1 for $\Lambda>0$ (see also \cite{bh002}) and Fig.\ 2 for $\Lambda<0$, replacing $\Lambda m$ and $\Lambda q$ with $\Lambda m^2$ and $\Lambda q^2$ respectively,
where Figs. 1-2 represent the five-dimensional cases.
The solid (red) curve in Fig.\ 1 corresponds to the {\it extremal} solutions of $r_+=r_-$, while the dashed (blue) curve
corresponds to the {\it charged Nariai} solutions of $r_+=r_c$.
The point where these two curves meet represents the ultracold solution of $r_+=r_-=r_c$. In the region surrounded by these two
curves and $q=0$ axis, three horizons exist at the same time.
For the case of $q=m$, obtained from $f'(r_+)=-f'(r_c)$,
there is the special solution with three horizons,
known as the {\it lukewarm} solution. The horizon positions in this solution are simply given by
\bea
&&r_c=\frac{\sqrt{3}}{2\sqrt{\Lambda}}
\Big(1+\sqrt{1-4m\sqrt{\frac{\Lambda
}{3}}}\Big),\,
r_+=\frac{\sqrt{3}}{2\sqrt{\Lambda}}
\Big(1-\sqrt{1-4m\sqrt{\frac{
\Lambda
}{3}}}\Big),\,
r_-=\frac{\sqrt{3}}{2\sqrt{\Lambda}}
   \Big(\sqrt{1+4m\sqrt{\frac{
\Lambda
}{3}}}-1\Big).
\label{luke}
\eea
Such a simple analytic expression for the lukewarm
solutions can not be found in other dimensions,
although the lukewarm solutions themselves can also exist.

For $\Lambda<0$, the phase diagram is essentially the same as the Fig. 2, replacing $|\Lambda|m$ and $|\Lambda| q$ with $|\Lambda| m^2$ and $|\Lambda| q^2$, respectively. The solid curve shows the extremal solution and below this curve a BH has both the outer and inner
horizons.

\subsection{The five-dimensional case}

In the case of $n=4$, the Eq.\ (\ref{metrics}) becomes
\bea
&&ds^2= -f(r) dt^2+\frac{dr^2}{f(r)}
   +r^2 d\Omega_{3}^2,
\quad
f(r)=1-\frac{\Lambda r^2}{6}
-\frac{2m}{r^2}
+\frac{q^2}{r^4},
\nonumber\\
&& F_{\chi\theta\phi}=\frac{\sqrt{6}q}{\kappa}\sin^2\chi \sin \theta,
\eea
where $d\Omega_{3}^2=d\chi^2+\sin^2\chi
\big(d\theta^2+\sin^2\theta d\phi^2\big)$ denotes the unit 3-sphere and we take $\epsilon_{\chi\theta\phi}=+1$.

For $\Lambda>0$, the existence of three horizons is ensured by
three curves on the $(\Lambda m,\Lambda q)$ plane, namely $q=0$ and
\bea
&&q\Lambda
=
q_{\rm ext}\Lambda:=
\frac{2}{\sqrt{3}}
\big(
(-2+3\Lambda m)+2(1-\Lambda m)^{\frac{3}{2}}
\big)^{\frac{1}{2}},\label{ext}
\\
&&
q\Lambda
=
q_{\rm cN}\Lambda
:=
\frac{2}{\sqrt{3}}
\big(
(-2+3\Lambda m)-2(1-\Lambda m)^{\frac{3}{2}}
\big)^{\frac{1}{2}},\label{cN}
\eea
where $q_{\rm ext}$ and $q_{\rm cN}$ correspond to the charges of the extremal and charged Nariai solutions, which are $r_+=r_-$ and $r_c=r_+$, respectively. The phase of a magnetically BH with a positive cosmological constant in five dimensions is shown in Fig. 1. $q_{\rm ext}$ and $q_{\rm cN}$ are shown by the solid (red) and dashed (blue) curves, respectively. There is no essential difference from the four-dimensional case. The point where these two curves meet, where $\Lambda m=1$ and $\Lambda q=\frac{2}{\sqrt{3}}$,
represents the ultracold solution, where
$r_+=r_-=r_c=\sqrt{\frac{2}{\Lambda}}$. There is also the lukewarm condition satisfying $f'(r_+)=-f'(r_c)$. As mentioned previously,
there is no simple analytic expression for the horizons as in four dimensions. Note that the ultracold solution corresponds to the common end point of the previous two curves. In the closed region surrounded by these curves, there are three nondegenerating horizons. Outside this region, there is only the cosmological horizon. The essential picture remains the same in other dimensions.

For $\Lambda<0$, the condition for existing two BH horizons is given by
\bea
q|\Lambda|
\leq
q_{\rm ext}|\Lambda|
:=
\frac{2}{\sqrt{3}}
\big(
2(1+|\Lambda| m)^{3/2}-(2+3|\Lambda|m)
\big)^{1/2}.
\eea
The phase of a magnetically charged BH in five dimensions is shown in Fig.\ 2. Note that there is no endpoint of the curve. The essential picture remains the same in other dimensions.

\section{Charged BH pair separated by a domain wall}

After giving BH solutions in the four and five dimensions,
we construct the solution representing a pair of charged BHs separated by a domain wall. We assume the case with the
$Z_2$ symmetry with respect to the domain wall.

\subsection{BH pair separated by a domain wall}

From now on, we work in the Euclidean section. We consider a pair of charged BHs separated by a domain wall. We focus on the Euclidean action involving the contribution of a boundary domain wall
\bea
\label{ck}
S_E&=&-\int_M d^{n+1} x\sqrt{g}
\left[
\frac{1}{2\kappa^2}\Big(R-2\Lambda\Big)
-\frac{1}{2 (n-1)! }
F_{\alpha_1 \cdots \alpha_{n-1}}F^{\alpha_1 \cdots \alpha_{n-1}}
\right] \nonumber \\
&+&\oint_{\partial M} d^n x\sqrt{q}\left[\sigma
+\frac{K-K_o}{\kappa^2}+ \frac{1}{(n-2)!} F^{\alpha_1 \cdots \alpha_{n-1}}n_{[\alpha_1} A_{\alpha_2 \cdots \alpha_{n-1}]}\right] ,
\eea
where $M$ and $\partial M$
represent the bulk spacetime and the domain wall, respectively.
Here, $\sigma$ denotes the tension of the domain wall. The last two terms denote the boundary term for gravity and the form field.

The magnetically charged BH solutions in the Euclidean section
are given by
\bea
&&ds_E^2=f(r) dt^2
+\frac{dr^2}{f(r)}
+r^2 d\Omega_{(n-1)}^2,
\quad
f(r)=1-\frac{2\Lambda r^2}{n(n-1)}
      -\frac{2m}{r^{n-2}}
      +\frac{q^2}{r^{2(n-2)}},
\eea
where the avoidance of the conical singularity at the outer horizon $r=r_+$ imposes the periodicity of $t$ coordinate to be
\bea
\beta
=\frac{4\pi}{|f'(r_{+})|}.\label{beta}
\eea
Here, $\beta$ is the inverse of Hawking temperature of BHs. Applying the cut-and-paste method or employing the instanton solutions mediating tunneling between the degenerate vacua, the solution representing an oppositely charged BH pair separated by a domain wall is constructed. Now we will employ the Israel junction condition \cite{israel}. In this framework, Einstein equations should be solved on either side of the wall. In the Euclidean section, a domain wall is moving along the trajectory characterized by the affine parameter $\tau$, $r=r(\tau)$. $\tau$ is normalized to satisfy $\frac{\dot r^2}{f}+f\dot t^2=1$, where ``dot'' means the derivative with respect to $\tau$. The equation of motion for $r(\tau)$ is given by $\dot{r}^2-V(r)=0$, where the effective potential is given by
\bea
V(r):=
     f(r)
      -\frac{\kappa^4 \sigma^2}{4(n-1)^2}r^2.
\eea
By redefining the parameters, the potential is given by
\bea
\label{red_pot}
V(r)=
 -A r^2
+1
- \frac{2m}{r^{n-2}}
+\frac{q^2}{r^{2(n-2)}},
\eea
where
\bea
\label{defA}
A:=\frac{2\Lambda}{n(n-1)}
           +\frac{\kappa^4 \sigma^2}{4(n-1)^2}.
\eea

From Eq. (\ref{defA}), in general we find
\bea
A\geq\frac{2\Lambda}{n(n-1)}.
\label{Acos}
\eea
On the other hand, for $A<0$ (and $\Lambda<0$),
\bea
|A|\leq \frac{2|\Lambda|}{n(n-1)}.\label{Acar}
\eea

We then regard $A$ as an input parameter, rather than the tension $\sigma$ itself. A stationary domain wall can exist at an equilibrium point $r=r_{\ast}$, which satisfies
\bea
V(r_{\ast})=\frac{dV}{dr}\Big|_{r=r_\ast}=0.
\label{wall_position}
\eea
The corresponding mass of the BH is denoted by $m_{\ast}$.
In the next subsections, we will obtain the explicit expressions for $r_{\ast}$ and $m_{\ast}$ in the four- and five-dimensional spacetimes. The induced metric on domain wall metric is given by
\bea
ds^2_{\rm ind}=d\tau^2
+r_{\ast}^2d\Omega_{n-1}^2.
\eea
For the domain wall at the equilibrium point, the Euclidean proper time on the wall has the periodicity of $\sqrt{f(r_{\ast})}\beta$,
which can be interpreted as the inverse of the temperature measured by an observer living on the domain wall. The system composed of a domain wall together with a BH pair is now in the thermal equilibrium.

In the semiclassical approximation, the production rate for a creation of a charged BH pair is given in Eq.\ (\ref{ck}). The boundary term for the form field vanishes for our solution.
From now on, we rewrite the Euclidean action useful for the evaluation. By employing the Israel junction conditions
$[K]=-\frac{n}{n-1}\kappa^2 \sigma$ and the Einstein equations,
the Euclidean action can be rewritten as
\bea
S_E&=&-\int_M d^{n+1} x\sqrt{g}
\left[
 \frac{2\Lambda}{(n-1)\kappa^2}
-\frac{n-2}{(n-1) (n-1)!}F_{(n-1)}^2
\right]
-\frac{1}{n-1}\oint_{\partial M} d^{n}x\sqrt{q}\sigma.
\label{ons}
\eea
Note that the Euclidean geometry is smooth at the horizon because of
our choice of $\beta$ as Eq.\ (\ref{beta}), and hence there is no contribution of the horizon to $S_E$. We therefore take only the boundary term of the domain wall into considerations.
Our strategy is to compute the Euclidean action for our system in four- and five- dimensional spacetimes and compare our results of the charged case with those of the neutral case with a same cosmological constant. After plugging $F_{(n-1)}^2=\frac{[(n-1)!]^2 q^2}{\kappa^2r^{2(n-1)}}$ into Eq.\ (\ref{ons}), which further reduces to
\bea
S_E&=&-\frac{2\Omega_{n-1}}{(n-1)\kappa^2}
\beta
 \Big[\frac{2 \Lambda}{n}\big(r_{\ast}^n-r_{+}^n\big)
      +(n-1)! q^2 \Big(\frac{1}{r_{\ast}^{n-2}}
               -\frac{1}{r_{+
}^{n-2}}
            \Big) \Big]
\nonumber \\
&-&\frac{
\sigma r_{\ast}^{n-1} \Omega_{n-1}}
{(n-1)}
 \beta
\sqrt{f (r_{\ast})}.
\eea
From Eq.\ (\ref{wall_position}) and Eq.\ (\ref{beta}),
we find
\bea
f(r_{\ast})=\Big(A-\frac{2\Lambda}{n(n-1)}\Big)r_{\ast}^2,\quad
\beta=\frac{2\pi r_+}{\frac{n m_\ast}
{r_+^{n-2}}-1-\frac{(n-1)q^2}{r_+^{2n-4}}},
\eea
respectively. Thus, we finally obtain
\bea
\label{ndim}
S_E&=&
-\frac{2\Omega_{n-1}}{(n-1)\kappa^2}
 \frac{2\pi r_+}
{\frac{n m_{\ast}}{r_+^{n-2}}
-1-
\frac{(n-1)q^2}{r_+^{2n-4}}}
\nonumber\\
&\times&
 \Big[
-\frac{2\Lambda}{n}r_+^n
+(n-1) A r_{\ast}^n
+(n-1)q^2
\Big(\frac{1}{r_{\ast}^{n-2}}
-\frac{1}{r_+^{n-2}}
\Big)
\Big],
\eea
where $m_{\ast}$ is the mass of the BH when the domain wall is placed at $r=r_{\ast}$. For later convenience, we introduce the dimensionless quantities measured in the unit of charge
\bea
\label{dimless}
&&\hr:=\frac{r}{q^{\frac{1}{n-2}}},\quad
\hA:=A q^{\frac{2}{n-2}},\quad
{\hat \Lambda}:= \Lambda q^{\frac{2}{n-2}},\quad
\m:=\frac{m}{q},\quad
{\hat \kappa}^2:=\frac{\kappa^2}{q^{\frac{n-1}{n-2}}},
\quad
{\hat \sigma}:= q^{\frac{n}{n-2}}\sigma.
\eea

\subsection{The four-dimensional case}

For the case of $n=3$, from Eq.\ (\ref{wall_position})
we obtain
\bea
\label{ackon}
&&-1+Ar_{\ast}^2+\frac{2m}{r_{\ast}}-\frac{q^2}{r_{\ast}^2}=0,
\quad
2Ar_{\ast}^2-\frac{2m}{r_{\ast}}+\frac{2q^2}{r_{\ast}^2}=0.
\eea
For $A>0$,
the solution to Eq.\ (\ref{ackon}) is given by
\bea
\hr_{\ast,\pm}=\left(\frac{1\pm \sqrt{1-12\hA}}{6\hA}\right)^{1/2},
\quad
\m_{\ast,\pm}
=\frac{\sqrt{2}}{6\sqrt{3\hA}}
\frac{1+12 \hA \pm \sqrt{1-12 \hA }}
     {\sqrt{1\pm \sqrt{1-12\hA }}},
\eea
The positions of $r_{\ast,+}$ and $r_{\ast,-}$ correspond to the local maximum and minimum of $V(r)$, respectively.

Thus, for $0\leq {\hat A}\leq \frac{1}{12}$, there are two roots,
which we call the (+)- and (-)-branch, respectively. For $\hA> \frac{1}{12}$, there is no root in both branches. We find $m_{\ast,\pm}^2> \frac{8}{9}q^2$. For a (+)-branch domain wall, $m_{\ast,+}^2>q^2$ for $0<\hA<\frac{1}{16}$. For a (-)-branch domain wall, $\frac{8}{9}q^2 \leq m_{\ast,-}^2<q^2$ for $0<\hA\leq \frac{1}{12}$. A special case is $\hA=\frac{1}{12}$, where
the degenerate root is given by $r_{\ast}=\sqrt{2}q$, resulting in
$m_{\ast}^2= \frac{8}{9}q^2=\frac{4}{9}r_{\ast}^2$. This is the case of a pair of the ultracold BHs $r_+=r_-=r_c$, where $m^2\Lambda=\frac{2}{9}$ and $q^2 \Lambda=\frac{1}{4}$.
Thus, from the definition of $A$, we find $\sigma=0$.
The other special case is a pair of the lukewarm BHs Eq. (\ref{luke}). This is obtained for $m_{\ast,\pm}^2=q^2$,
namely, for $\hA=\frac{1}{16}$ of the (+)-branch and
for $\hA=0$ of the (-)-branch. In the limit of $q\to 0$,
\bea
r_{\ast,+} \to \frac{1}{\sqrt{3A}},\quad
r_{\ast,-}\to 0.
\eea
Thus, only for the (+)-branch can the domain wall exist.
The corresponding mass of the BH is given by
\bea
m_{\ast,+}\to \frac{1}{3\sqrt{3A}}.
\eea

Until now, we have not taken the existence of the outer horizon into consideration. However, it is straightforward to confirm that
there is no real solution of the outer horizon for the $(+)$-branch of $A\leq\frac{\Lambda}{3}$ and for the $(-)$-branch of $A\geq\frac{\Lambda}{3}$. It means that for the $(-)$-branch
no horizon can be formed for $A\geq\frac{\Lambda}{3}$ obtained
from Eq.\ (\ref{Acos}) with $n=3$. For the special lukewarm solution, $r_{\ast,+}=2q=2m_{\ast}$, and then $r_c>r_{\ast}>r_+$,
since $r_c>2q$ and $r_+<2q$, which ensures that the (+)-branch domain wall is always located between the cosmological and outer horizons.
On the other hand, however, the $(-)$-branch domain wall for the lukewarm solution may exist for $A\to 0$ with $q\neq 0$. But it contradicts the condition $A>\frac{\Lambda}{3}$,
and hence no $(-)$-branch BH pair is formed.

For $A<0$, the solution to Eq.\ (\ref{ackon}) is given by
\bea
\hr_{\ast}=
\left(\frac{-1+ \sqrt{1+12|\hA|}}
{6|\hA|}\right)^{1/2},\quad
\m_{\ast}
=\frac{\sqrt{2}}{6\sqrt{3|\hA|}}
\frac{-1+12 |\hA|+ \sqrt{1+12|\hA| }}
     {\sqrt{-1+ \sqrt{1+12|\hA| }}}.
\eea
The position of $r_{\ast}$ corresponds to the local minimum of $V(r)$. For $|\hA|\to 0$, $m_{\ast} \to q$ and for larger $|A|$, $\m_{\ast}^2$ is increasing. In the limit of $q\to 0$,
$r_{\ast}\to 0$.

However, it is straightforward to confirm that for $A<0$
no horizon is formed for $|A|\leq \frac{|\Lambda|}{3}$.
But from Eq. (\ref{Acar}) with $n=3$, we must have
$|A|\leq \frac{|\Lambda|}{3}$. Therefore, the horizon cannot be formed for $A<0$.

The phase diagram for a domain wall in four dimensions
is quite similar to Fig. 3 which is for the five-dimensional case
discussed in the next subsection.
\begin{figure}
   \begin{center}
    \includegraphics[scale=.80]{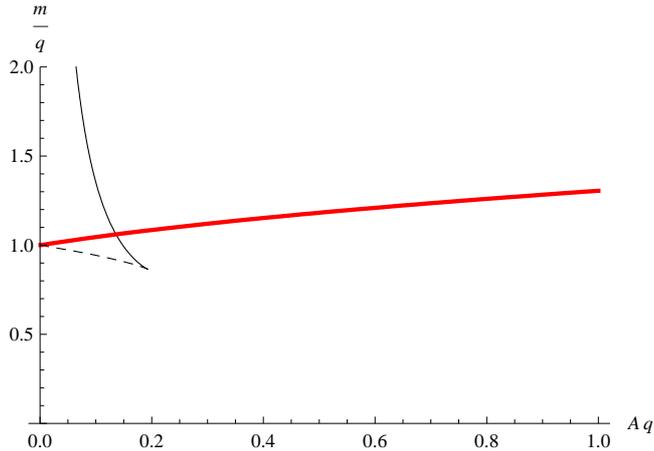}
        \caption{
The mass of the BH $\hat m_{\ast}$ is shown
as the function of $|\hA|$
in the five-dimensional case.
The solid, dashed and thick (red) curves represent
$\m_{\ast,+}$, $\m_{\ast,-}$ (for $A>0$)
and $\m_{\ast}$ (for $A<0$)
as functions of $|\hat A| $, respectively.
For the case of the four dimensions,
we obtain essentially the same picture.
There is no endpoint for the curve of $A<0$.
}
   \end{center} \label{fig3}
\end{figure}

\subsection{The five-dimensional case}

Along the similar arguments to the case of four dimensions,
we discuss the properties of a domain wall in the five-dimensional spacetime. For $A>0$, the position of the domain is given by either of the following two branches
\bea
&&
{\hat r}_{\ast,+}^2
=\frac{1}{6\hat A}
\Big[
1+C({\hat A})^{-1/3}
+C({\hat A})^{1/3}
\Big],
\nonumber\\
&&
{\hat r}_{\ast,-}^2
=\frac{1}{12\hat A}
\Big[
2+\big(-1+\sqrt{3}i\big)C({\hat A})^{-1/3}
+(-1-\sqrt{3}i)C({\hat A})^{1/3}
\Big].
\eea
The positions of $r_{\ast,+}$ and $r_{\ast,-}$ correspond to the local maximum and minimum of $V(r)$, respectively. For $A<0$,
\bea
{\hat r}_{\ast}^{2}
=\frac{1}{6|\hat A|}
\Big[
-1+D(|{\hat A}|)^{-1/3}
+D(|{\hat A}|)^{1/3}
\Big].
\eea
Here, the position of $r_{\ast}$ corresponds to the local minimum of $V(r)$. We defined
\bea
&&C(\hat A):=1-54\hat A^2
+6\sqrt{-3{\hat A}^2+81{\hat A}^4},
\nonumber\\
&&D(|\hat A|)
:=-1+54 |{\hat A}|^2
+6\sqrt{-3|{\hat A}|^2+81|{\hat A}|^4},
\eea
and the expression for $r_{\ast,-}$ is real. Thus, we find ${\hat A}\leq \frac{1}{3\sqrt{3}}$. In the limit of $q\to 0$ for $A>0$,
only for the (+)-branch there is the domain wall solution
$r_{\ast,+}^2\to\frac{1}{2A}$. For $A<0$, there is no domain wall solution of $q\to 0$.

For $A>0$, the corresponding BH mass is given by
\bea
{\hat m}_{\ast,+}({\hat A})
&=&\frac{1}{24\hat A}
\big(
1-{\hat C}^{-2/3}({\hat A})
+2{\hat C}^{-1/3}({\hat A})
+2{\hat C}^{1/3}({\hat A})
-{\hat C}^{2/3}({\hat A})
\big),
\nonumber\\
{\hat m}_{\ast,-}({\hat A})
&=&\frac{1}{48\hat A C({\hat A})^{2/3}}
\times\Big\{
\big(-1-i\sqrt{3}\big)
\big(1+12\sqrt{-3{\hat A}^2+81{\hat A}^4}\big)
-(1-i\sqrt{3})C({\hat A})^{1/3}
\nonumber\\
&+&2\big[
C({\hat A})^{1/3}
\big(3(1-i\sqrt{3})\sqrt{-3{\hat A}^2+81{\hat A}^4}
   +C({\hat A})^{1/3}
\big)
\nonumber\\
&+&27{\hat A}^2
\big(2(1+i\sqrt{3})
-C({\hat A})^{1/3}(1-i\sqrt{3})
\big)
\big]
\Big\}.
\eea
Similarly for $A<0$, the mass becomes
\bea
{\hat m}_{\ast}({\hat A})
=\frac{1}{24|{\hat A}|}
\big(
-1+{ D}^{-2/3}({\hat A})
+2{ D}^{-1/3}({\hat A})
+2{ D}^{1/3}({\hat A})
+{ D}^{2/3}({\hat A})
\big).
\eea
In Fig.\ 3, we have shown the phase of a domain wall in five dimensions. For $\hat A=\frac{1}{3\sqrt{3}}$, mass in the both branch becomes the degenerate value $m_{\ast,\pm}=\frac{\sqrt{3}}{2}q$,
which corresponds to the ultracold BHs. Then, from Eq. (\ref{defA}) for $n=4$ with $ \hat\Lambda=\frac{2}{\sqrt{3}}$ of the ultracold case, it turns out that the corresponding domain wall is tensionless. In the limit of $q\to 0$, the mass of the domain wall for the (+)-branch becomes $m_{\ast,+}\to \frac{1}{8A}$.

As in the four-dimensional case, however, it is straightforward to confirm that there is no outer horizon for the $(+)$-branch of $A\leq\frac{\Lambda}{6}$ and of for the $(-)$-branch of $A\geq\frac{\Lambda}{6}$. Therefore, for the $(-)$-branch no horizon can be formed since $A\geq\frac{\Lambda}{6}$ from Eq.\ (\ref{Acos}) with $n=4$. Similarly, it is straightforward to confirm that for $A<0$ no horizon can exist for $|A|\leq \frac{|\Lambda|}{6}$.
From Eq.\ (\ref{Acar}) with $n=4$, we must have $|A|\leq \frac{\Lambda}{6}$. Therefore, there is no BH pair for $A<0$.


\section{Pair creation rates}

In the semiclassical approximation, the formula of the production rate without a prefactor in the bounce approach is equivalent to the formula in the quantum cosmological approach \cite{bcha}. In the bounce approach, the production rate for a neutral BH pair in the background spacetime with a domain wall is given by $\Gamma_1 = A  e^{-[S_E(nbls)-S_E(bg)]}$, where $S_E(nbls)$ and $S_E(bg)$ mean the Euclidean action of neutral BHs with a domain wall and the action of the wall, respectively. The charged case also has the same form. Thus the ratio of two production rates is written by $\Gamma = \Gamma_2/\Gamma_1= e^{-[S_E(cbls)-S_E(nbls)]}$, where $S_E(cbls)$ means the Euclidean action of charged BHs with a domain wall. In the quantum cosmological approach, the ratio of two probabilities is given by $\Gamma=P_2/P_1 =e^{-[S_E(cbls)-S_E(nbls)]}$, where $P_2=e^{-S_E(cbls)}$ and $P_1=e^{-S_E(nbls)}$ mean the probability of charged BHs with a domain wall from nothing and of  neutral BHs with a domain wall from nothing. This method does not need the interpolating instanton solution. In the approximation, the half of the Euclidean action is used in the wave function. The probability is related to the real part of the Euclidean action in the quantum cosmological approach. This action is equivalent to the action, $S_E$, in the present work.

In this section, we compute the Euclidean action for each solution representing a BH pair separated by a domain wall in four- and five-dimensional spacetimes by employing the quantum cosmological approach. As we have seen in the previous section, the outer horizon of BHs can be formed only for the $(+)$-branch of the $A>0$ case. Thus, in this section we will focus on these cases. We then evaluate the nucleation rates.

\subsection{The four-dimensional case}

\subsubsection{For a pair of neutral BHs}

For a pair of the four-dimensional neutral BHs, a domain wall solution can exist only for $A>0$, with $m_{\ast}=\frac{1}{3\sqrt{3A}}$ and
$r_{\ast}=3m_{\ast}=\frac{1}{\sqrt{3A}}$. Thus, we have to focus on the case of $A>0$. From Eq.\ (\ref{ndim}) for $n=3$ with we obtain
\bea
S_E(nbls)&=&
-\frac{16\pi^2 r_+}{3\kappa^2|1-\Lambda r_+^2|}
\Big(-2\Lambda r_{+}^3+\frac{2}{\sqrt{3}A^{1/2}}\Big).
\eea
From now on, we discuss the cases of $\Lambda>0$ and of $\Lambda<0$,
separately. For $\Lambda>0$, we obtain
\bea
r_+=\frac{2}{\Lambda^{1/2}}
\cos\big(\frac{1}{3}(\pi+\theta)\big),
\eea
where
$\cos\theta:=x^{\frac{1}{2}}$ and $\sin\theta=\sqrt{1-x}$
with $x:=\frac{\Lambda}{3A}=\frac{1}{3\tilde A}$.
Here a quantity with ``tilde'' is dimensionless, and
normalized by an appropriate power of $\Lambda$. Since $0<x<1$,
$0<\cos\big(\frac{1}{3}(\pi+\theta)\big)<\frac{1}{2}$.
Finally, the Euclidean action is obtained by
\bea
S_E(nbls)= -\frac{64\pi^2}{3\kappa^2\Lambda}F_{4,Sch}(\tilde A),\quad
F_{4,Sch}(\tilde A):=3\cos^2\big(\frac{1}{3}(\pi+\theta)\big).
\label{measure4}
\eea
To derive it, we have employed the formula $\cos3\alpha=4\cos^3\alpha-3\cos\alpha$. It is easy to confirm that
\bea
 {F}_{4,Sch}=\frac{3r_+^2\Lambda}{4},
\eea
which leads to
\bea
S_{E}(nbls)=-\frac{16\pi^2 r_+^2}{\kappa^2}
=-\frac{2\pi r_+^2}{G}
=-\frac{{\cal A} }{4G},\label{entropy4}
\eea
where ${\cal A}=2\times (4\pi r_+^2)$ and $G=\frac{\kappa^2}{8\pi }$
represent the area of horizons (the factor $2$ means two sides) and the effective gravitational coupling. Thus, the absolute value of the Euclidean action is equal to the entropy of created BHs, following the area law. This result is somewhat natural since the domain wall with a pair of BHs is now in the thermal equilibrium. As we will see below, the same relation also holds for a charged BH pair with a domain wall, as well as in the five-dimensional cases. Note that the cosmological horizon is always out of our system and the thermal contribution from it is absent.

Similarly, for $\Lambda<0$ we obtain
\bea
r_+
=\frac{2}{\Lambda^{1/2}}
\sinh\big(\frac{1}{3}\Theta\big),
\eea
where
$\sinh\Theta:=y^{\frac{1}{2}}$
with $y:=\frac{|\Lambda|}{3A}=\frac{1}{3\tilde A}$.
Here a quantity with tilde is dimensionless, and
normalized by an appropriate power of $|\Lambda|$.
Finally, the Euclidean action is obtained by
\bea
S_E(nbls)=
-\frac{64\pi^2}{3\kappa^2|\Lambda|}
{\cal F}_{4,Sch}(\tilde A),\quad
{\cal F}_{4,Sch}(\tilde A):=
3\sinh^2\big(\frac{1}{3}\Theta\big).
\label{measure4_2}
\eea
To derive it, we employed the formula $\sinh(3\alpha)
=4\sinh^3\alpha+3\sinh\alpha$. It is easy to confirm that
\bea
 {\cal  F}_{4,Sch}=\frac{3r_+^2|\Lambda|}{4},
\eea
which leads to Eq.\ (\ref{entropy4}). The absolute value of the Euclidean action is equal to the entropy of created BHs,
following the area law.

\subsubsection{For a pair of lukewarm BHs}

For a charged BH pair, the only analytically tractable case
is the case of a pair of lukewarm BHs in four dimensions, in which
$r_{\ast}=2m_{\ast}=\frac{1}{2\sqrt{A}}$,
are satisfied. In this solution, the position of the outer horizon is given by
\bea
r_+=\frac{\sqrt{3}}{2\sqrt{\Lambda}}
 \Big(1-\sqrt{1-\Big(\frac{\Lambda}{3A}\Big)^{1/2}}\Big).
\eea
From Eq. (\ref{ndim}) for $n=3$ with
\bea
\beta=\frac{2\pi\sqrt{3}}{\sqrt{\Lambda}}
\frac{1}{\sqrt{1-\Big(\frac{1}{3\tilde A}\Big)^{1/2}}},
\eea
the Euclidean action becomes
\bea
S_E(cbls)&=&-\frac{64\pi^2}{3\kappa^2\Lambda}
F_{4, luke}(\tilde A),\quad
 F_{4, luke}(\tilde A)
:=\frac{9}{16}(1-\sqrt{1-\big(\frac{1}{3\tilde A}\big)^{1/2}})^2 .
\label{measure_luke}
\eea
For a given $\Lambda$, it is straightforward to see $F_{4,luke}(\tilde A)<F_{4,Sch}(\tilde A)$ for any $\tilde A$ (See Fig. 4), where $F_{4,Sch}$ is given in Eq.\ (\ref{measure4}).
Thus, the production probability $P_2=e^{-S_E(cbls)}$
of a lukewarm BH pair is always smaller than that of the neutral case. It is easy to find
\bea
  F_{4, luke}=\frac{3r_+^2\Lambda}{4},
\eea
which leads to
\bea
S_{E}(cbls)=-\frac{16\pi^2 r_+^2}{\kappa^2}
=-\frac{2\pi r_+^2}{G}
=-\frac{{\cal A} }{4G}.\label{entropy05}
\eea
As for the neutral case,
the absolute value of the Euclidean action is equal
to the entropy of created BHs, following the area law.


\subsubsection{For a pair of general charged BHs}

We focus on the case of the $(+)$-branch domain wall with
$A>0$. Substituting $n=3$ into Eq. (\ref{ndim}),
\bea
S_E(cbls)=-\frac{16\pi^2 r_+}
    {\kappa^2\big(\frac{3m_\ast}{r_+}-1-\frac{2q^2}{r_+^2}\big)}
 \Big[
-\frac{1}{3}r_+^3 \Lambda
+r_{\ast}^3 A
+q^2
\Big(\frac{1}{r_\ast}
    -\frac{1}{r_+}
\Big) \Big].
\eea
For $\Lambda>0$, for comparison with the case of a neutral BH pair,
it is useful to rewrite the Euclidean action
\bea
S_{E}(cbls)=-\frac{64\pi^2}{3\kappa^2\Lambda} F_{4},
\eea
where
\bea
F_4({\tilde A},{\tilde q}):=\frac{3{\tilde r}_+}{4}
\frac{1}
      {\frac{3{\tilde m}_{\ast}}{{\tilde r}_+}
      -\frac{2{\tilde q}^2}{{\tilde r}_+^2}-1}
\Big[
{\tilde A}{\tilde r}_\ast^3
-\frac{{\tilde r}_+^3}{3}
+{\tilde q}^2
\Big(
 \frac{1}{{\tilde r}_{\ast}}
-\frac{1}{{\tilde r}_+}
\Big)
\Big].
\eea
Here a quantity with tilde is dimensionless, and
normalized by an appropriate power of $\Lambda$.
Note that $F_4(\tilde A,0)=F_{4,Sch}(\tilde A) $
defined in Eq. (\ref{measure4}), and $F_4(\tilde A,\frac{1}{4\tilde A^{1/2}}) = F_{4,luke}(\tilde A)$ defined in Eq. (\ref{measure_luke}). We then numerically evaluate the function $F_{4}(\tilde A, q)$ for the general cases. In Fig.\ 4, the comparison of $F_4(\tilde A,\tilde q)$, $F_{4,Sch}(\tilde A)$ and $F_{4,luke}(\tilde A)$ is shown. Note that for this choice of parameters, $\frac{1}{3}\approx 0.333<\tilde A<1$. In Fig.\ 5, $F_4$ is shown as a function of $\tilde q$ for a fixed $\tilde A$. We find that $F_4(\tilde A,\tilde q)<F_{4,Sch}(\tilde A)$. Therefore, the pair production rate for a charged BH is always smaller than that of the neutral case. We then numerically confirmed
\bea
 {F}_4=\frac{3r_+^2\Lambda}{4},
\eea
which leads to Eq.\ (\ref{entropy05}). As for the neutral case,
the absolute value of the Euclidean action is equal
to the entropy of created BHs, following the area law.
Then, the decreasing pair creation rate for the charged case
can be understood in terms of the decreasing area of the outer horizon for increasing charge with a fixed cosmological constant.
\begin{figure}
   \begin{center}
    \includegraphics[scale=.80]{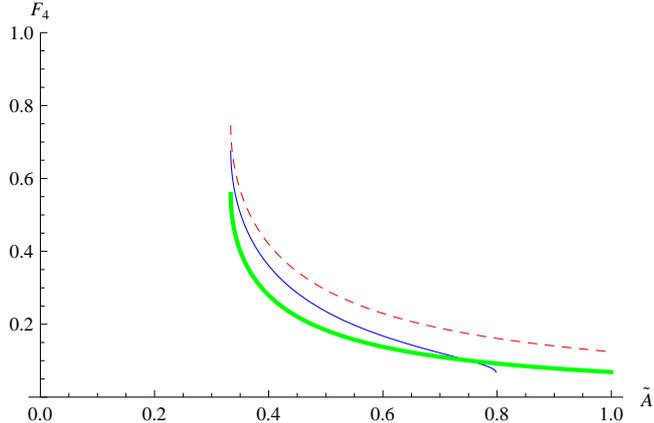}
        \caption{
$F_4 (\tilde A,\tilde q)$, $F_{4,Sch}(\tilde A)$
and $F_{4, luke}(\tilde A)$
are shown by the solid (blue), dashed (red)
and thick (green) curves,
respectively,
as the function of $\tilde A$ for
$\tilde q=\frac{1}{2\sqrt{3}}\approx 0.289$.}
   \end{center} \label{fig4}
\end{figure}
\begin{figure}
   \begin{center}
    \includegraphics[scale=.80]{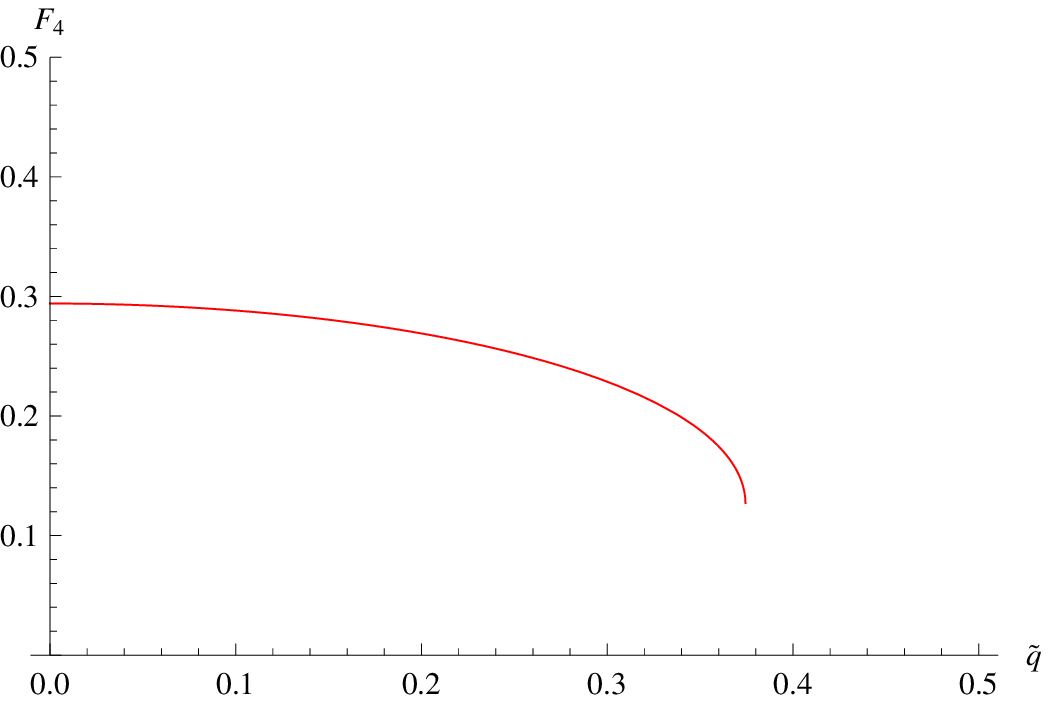}
        \caption{
${F}_4 (\tilde A,\tilde q)$
is shown by the solid (red)
curve
as the function of $\tilde q$ for
$\tilde A=0.5$.
}
   \end{center}\label{fig5}
\end{figure}

Similarly for $\Lambda<0$,
\bea
S_{E}(cbls)=-\frac{64\pi^2}{3\kappa^2|\Lambda|}
{\cal F}_{4},
\eea
where
\bea
{\cal F}_4({\tilde A}, {\tilde q}):=\frac{3{\tilde r}_+}{4}
\frac{1}
      {\frac{3{\tilde m}_{\ast}}{{\tilde r}_+}
      -\frac{2{\tilde q}^2}{{\tilde r}_+^2}-1}
\Big[
{\tilde A}{\tilde r}_\ast^3
+\frac{{\tilde r}_+^3}{3}
+{\tilde q}^2
\Big(
 \frac{1}{{\tilde r}_{\ast}}
-\frac{1}{{\tilde r}_+}
\Big)
\Big],
\eea
Here a quantity with tilde is dimensionless, and
normalized by an appropriate power of $|\Lambda|$.
Note that ${\cal F}_4(\tilde A,0)={\cal F}_{4,Sch}(\tilde A) $
defined in Eq.\ (\ref{measure4_2}). In Fig.\ 6,
the comparison of ${\cal F}_4(\tilde A,\tilde q)$ and
${\cal F}_{4,Sch}(\tilde A)$ is shown. In Fig.\ 7, ${\cal F}_4$ is shown as a function of $\tilde q$ for a fixed $\tilde A$.
Therefore, the pair production rate for a charged BH is smaller than
that of the neutral case. We then numerically confirmed
\bea
 {\cal F}_4=\frac{3r_+^2|\Lambda|}{4},
\eea
which leads to Eq. (\ref{entropy05}).
The absolute value of the Euclidean action is equal
to the entropy of created BHs, following the area law.
Similarly, the decreasing pair creation rate for the charged case
can be understood in terms of the decreasing area of the outer horizon for increasing charge with a fixed cosmological constant.
\begin{figure}
   \begin{center}
    \includegraphics[scale=.80]{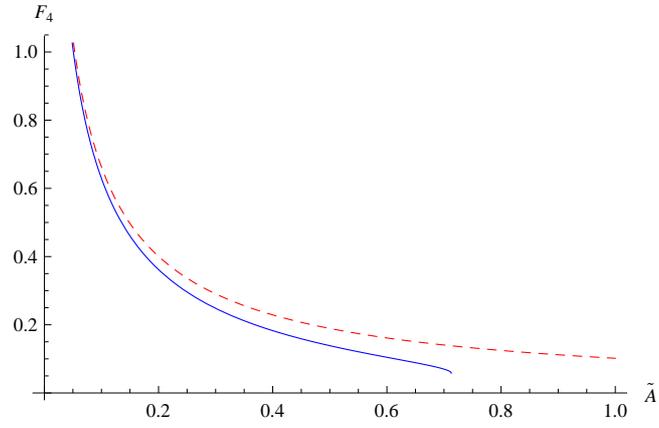}
        \caption{
${\cal F}_4 (\tilde A,\tilde q)$ and ${\cal F}_{4,Sch}(\tilde A)$
are shown by the solid (blue) and dashed (red)
curves, respectively,
as the function of $\tilde A$ for
$\tilde q=\frac{1}{2\sqrt{3}}\approx 0.289$.
}
   \end{center}\label{fig6}
\end{figure}
\begin{figure}
   \begin{center}
    \includegraphics[scale=.80]{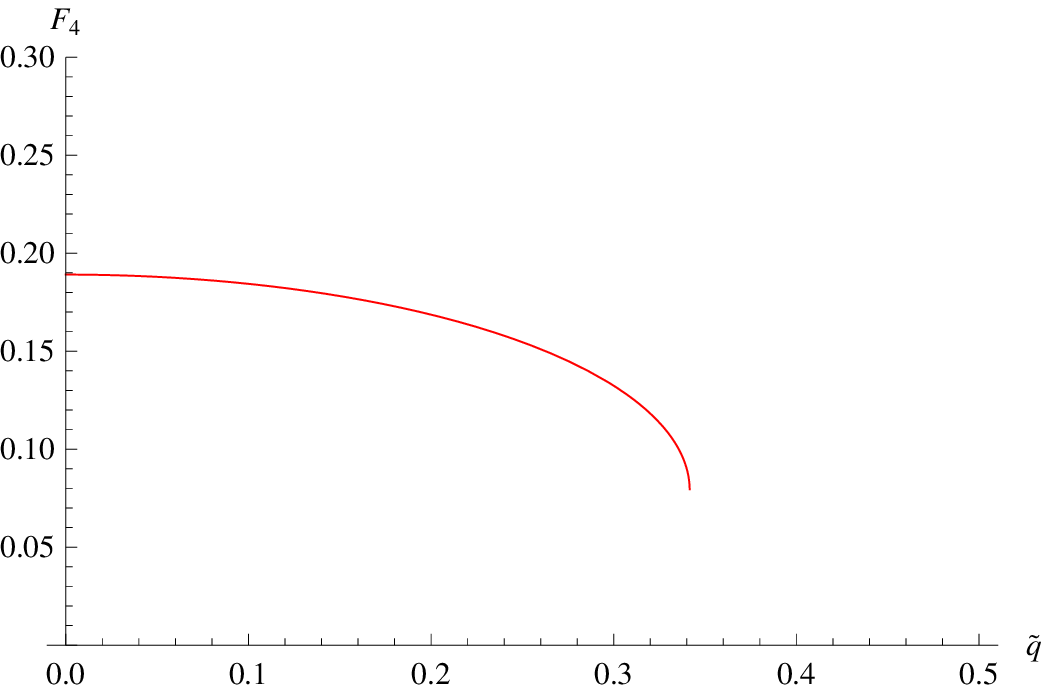}
        \caption{
${\cal F}_4 (\tilde A,\tilde q)$
is shown by the solid (red)
curve
as the function of $\tilde q$ for
$\tilde A=0.5$.
}
   \end{center}\label{fig7}
\end{figure}


\subsection{The five-dimensional case}

In this subsection, we repeat the similar computations
for the five-dimensional spacetime.

\subsubsection{For a pair of neutral BHs}

Similarly to the case of four dimensions, in the case of the neutral BHs $q=0$, the domain wall solution exists only for $A>0$, with $m_{\ast}=\frac{1}{8A}$ and
$r_{\ast}=\frac{1}{\sqrt{2A}}=2m_{\ast}^{\frac{1}{2}}$.
For $\Lambda>0$ and $\Lambda<0$, the BH horizon is located at
\bea
r_+^2=\frac{3}{\Lambda}\Big(1-\sqrt{1-\frac{\Lambda}{6A}}\Big),
\quad
\frac{3}{|\Lambda|}\Big(\sqrt{1+\frac{|\Lambda|}{6A}}-1\Big),
\eea
respectively. From Eq.\ (\ref{ndim}) for $n=4$, for $\Lambda>0$ we obtain
\bea
S_E(nbls)&=&
-\frac{(2 \sqrt{3})^3\pi^3}{\kappa^2\Lambda^{3/2}}
F_{5,Sch},
\quad
F_{5,Sch}(\tilde A)
:=\Big(1-\sqrt{1-\frac{1}{6\tilde A}}\Big)^{3/2},
\label{measure5}
\eea
where $\tilde A:=\frac{A}{\Lambda}$,
while for $\Lambda <0$ we obtain
\bea
S_E(nbls)&=&
-\frac{(2 \sqrt{3})^3\pi^3}{\kappa^2|\Lambda|^{3/2}}
{\cal F}_{5,Sch},\quad
{\cal F}_{5,Sch}(\tilde A)
:=\Big(\sqrt{1+\frac{1}{6\tilde A}}-1\Big)^{3/2},
\label{measure5_2}
\eea
where $\tilde A:=\frac{A}{|\Lambda|}$.
It is easy to confirm that
\bea
S_E(nbls)=-\frac{{\cal A}}{4G},\label{entropy5}
\eea
irrespective of the sign of $\Lambda$, where
${\cal A}=2\times (2\pi^2 r_+^3)$ is the area of black horizon.
As for the four-dimensional case, the absolute value of the Euclidean action is equal to the entropy of created BHs, following the area law. The result agrees with that for a bulk BH pair of Ref. \cite{gasa}.

\subsubsection{For a pair of general charged BHs}

We focus on the case of the $(+)$-branch domain wall with
$A>0$. Substituting $n=4$ into Eq.\ (\ref{ndim}),
the Euclidean action reduces to
\bea
S_E(cbls)&=&
-\frac{8\pi^3 r_+}
    {\kappa^2\big(\frac{4 m_{\ast}}{r_+^2}-1-\frac{3q^2}{r_+^4}\big)}
 \Big[
-\frac{1}{2}r_+^3{\Lambda}
+3r_{\ast}^3A
+6q^2\Big(\frac{1}{r_{\ast}^2}
-\frac{1}{r_+^2}\Big)
\Big].
\eea
For $\Lambda>0$,
\bea
S_{E}(cbls)=-\frac{(2\sqrt{3}\pi)^3}{\kappa^2\Lambda^{\frac{3}{2}}} F_{5},
\eea
where
\bea
F_5(\tilde A,\tilde q):=\frac{{\tilde r}_+}{9\sqrt{3}}
\frac{1}
      {\frac{4{\tilde m}_{\ast}}{{\tilde r}_+^2}
      -\frac{3{\tilde q}^2}{{\tilde r}_+^4}-1}
\Big[
3{\tilde A}{\tilde r}_\ast^4
-\frac{{\tilde r}_+^4}{2}
+6{\tilde q}^2
\Big(
 \frac{1}{{\tilde r}_{\ast}^2}
-\frac{1}{{\tilde r}_+^2}
\Big)
\Big].
\eea
Here a quantity with tilde is dimensionless,
and normalized by an appropriate power of $\Lambda$.
Note that $F_5(\tilde A,0)=F_{5,Sch}(\tilde A) $
defined in Eq.\ (\ref{measure5}). In Fig.\ 8,
the comparison of $F_5(\tilde A,\tilde q)$ and
$F_{5,Sch}(\tilde A)$ is shown. In Fig. 9, $F_5$ is shown as a function of $\tilde q$ for a fixed $\tilde A$. We find that $F_5(\tilde A,\tilde q)<F_{5,Sch}(\tilde A)$. Therefore, as in the four-dimensional case, the pair production rate for a charged BH is always smaller than that of the neutral case. We then numerically confirmed
\bea
 {F}_5=\frac{r_+^3\Lambda^{\frac{3}{2}}}{(\sqrt{3})^3},
\eea
which leads to
\bea
S_{E}(cbls)=-\frac{(2\pi r_+)^3}{\kappa^2}
=-\frac{\pi^2 r_+^3}{G}
=-\frac{{\cal A} }{4G},\label{entropy52}
\eea
where ${\cal A}=2\times (2\pi^2 r_+^3)$ and $G=\frac{\kappa^2}{8\pi }$
represent the area of horizons (the factor $2$ means two sides).
The absolute value of the Euclidean action is equal to the entropy of created BHs, following the area law. Then, the decreasing pair creation rate for the charged case can be understood
in terms of the decreasing area of the outer horizon
for the increasing charge with a fixed cosmological constant.
\begin{figure}
   \begin{center}
    \includegraphics[scale=.80]{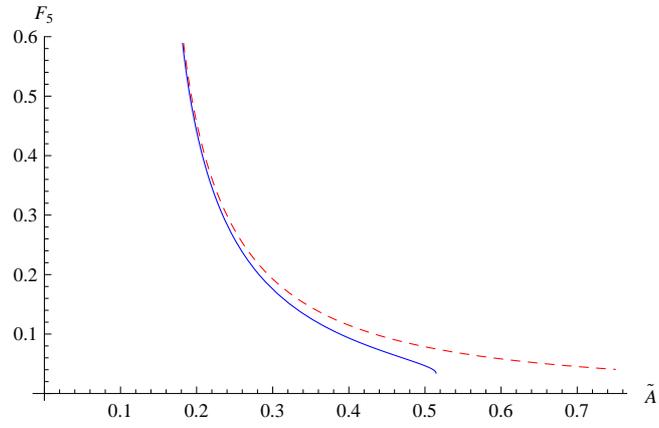}
        \caption{
$F_5 (\tilde A,\tilde q)$,
and $F_{5,Sch}(\tilde A)$
are shown by the solid (blue) and dashed (red)
curves,
respectively,
as the function of $\tilde A$ for
$\tilde q=0.3$.
}
   \end{center}\label{fig8}
\end{figure}
\begin{figure}
   \begin{center}
    \includegraphics[scale=.80]{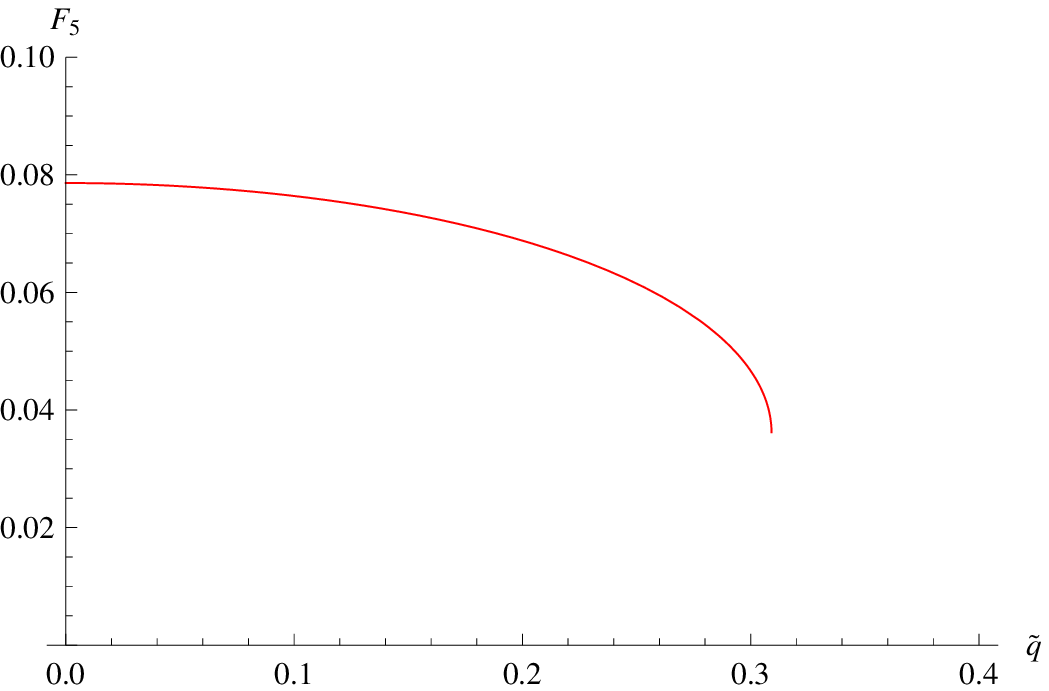}
        \caption{
${F}_5 (\tilde A,\tilde q)$
is shown by the solid (red)
curve
as the function of $\tilde q$ for
$\tilde A=0.5$.
}
   \end{center}\label{fig9}
\end{figure}

For $\Lambda<0$, we obtain
\bea
S_{E}(cbls)=-\frac{(2\sqrt{3}\pi)^3}{\kappa^2|\Lambda|^{\frac{3}{2}}}
{\cal F}_{5},
\eea
where
\bea
{\cal F}_5
(\tilde A,\tilde q):=\frac{{\tilde r}_+}{9\sqrt{3}}
\frac{1}
      {\frac{4{\tilde m}_{\ast}}{{\tilde r}_+^2}
      -\frac{3{\tilde q}^2}{{\tilde r}_+^4}-1}
\Big[
3{\tilde A}{\tilde r}_\ast^4
+\frac{{\tilde r}_+^4}{2}
+6{\tilde q}^2
\Big(
 \frac{1}{{\tilde r}_{\ast}^2}
-\frac{1}{{\tilde r}_+^2}
\Big)
\Big].
\eea
Here a quantity with tilde is dimensionless, and
normalized by an appropriate power of $|\Lambda|$.
Note that ${\cal F}_5(\tilde A,0)={\cal F}_{5,Sch}(\tilde A) $
defined in Eq.\ (\ref{measure5_2}). In Fig.\ 10,
the comparison of ${\cal F}_4(\tilde A,\tilde q)$ and
${\cal F}_{5,Sch}(\tilde A)$ is shown. In Fig.\ 11, ${\cal F}_5$ is shown as a function of $\tilde q$ for a fixed $\tilde A$.
We find that ${\cal F}_5(\tilde A,\tilde q)<{\cal F}_{5,Sch}(\tilde A)$. Therefore, the production rate for a charged BH pair is smaller than that of the neutral case. We then numerically confirmed
\bea
 {\cal F}_5=\frac{r_+^3|\Lambda|^{\frac{3}{2}}}{(\sqrt{3})^3},
\eea
which leads to Eq. (\ref{entropy52}). As for the neutral case,
the absolute value of the Euclidean action is equal to the entropy of created BHs, following the area law. Similarly, the decreasing pair creation rate for the charged case can be understood in terms of the decreasing area of the outer horizon for increasing charge
with a fixed cosmological constant.
\begin{figure}
   \begin{center}
    \includegraphics[scale=.80]{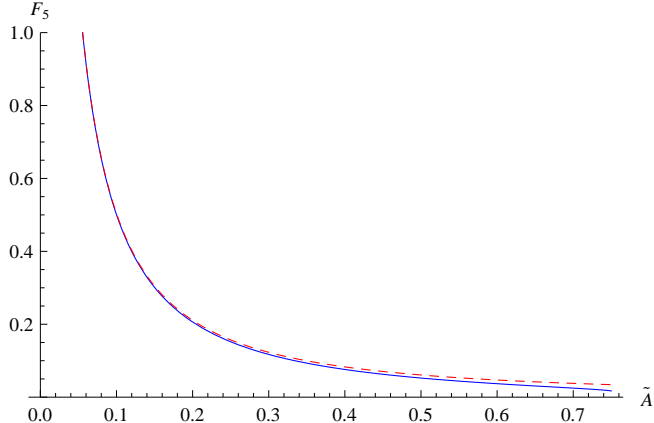}
        \caption{
${\cal F}_5 (\tilde A,\tilde q)$ and ${\cal F}_{5,Sch}(\tilde A)$
are shown by the solid (blue) and dashed (red) curves,
respectively,
as the function of $\tilde A$ for
$\tilde q=\frac{1}{3\sqrt{3}}\approx 0.192$.
}
   \end{center}\label{fig10}
\end{figure}
\begin{figure}
   \begin{center}
    \includegraphics[scale=.80]{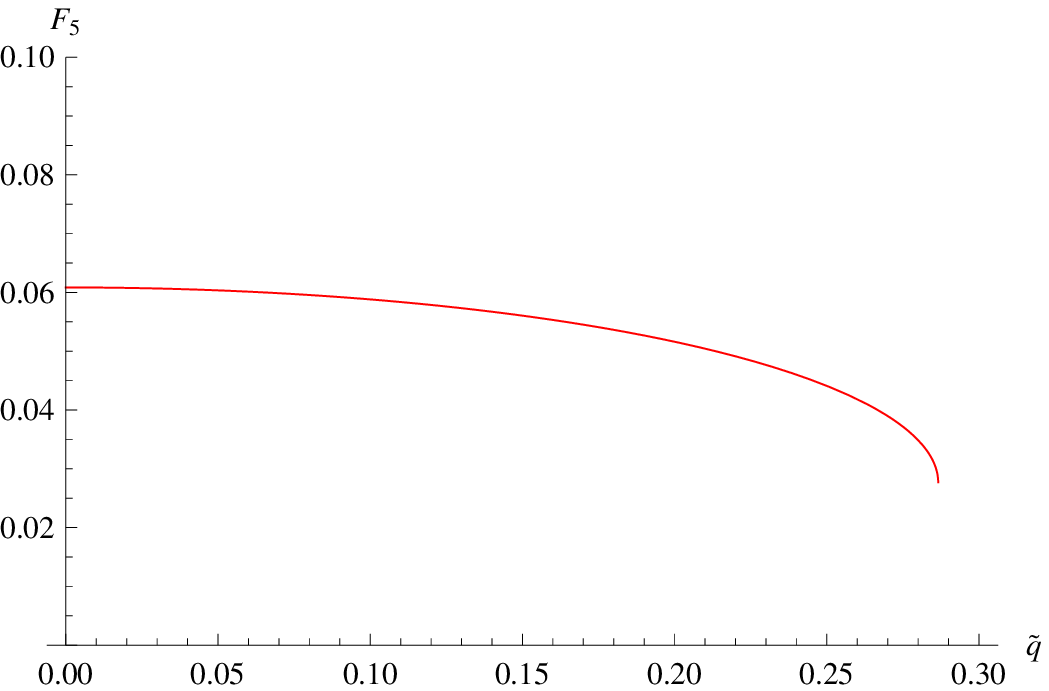}
        \caption{
${\cal F}_5 (\tilde A,\tilde q)$
is shown by the solid (red)
curve
as the function of $\tilde q$ for
$\tilde A=0.5$.
}
   \end{center}\label{fig11}
\end{figure}

\subsection{Application to the braneworld cosmology}

After the nucleation, the domain wall may evolve in the radial direction of the bulk spacetime. The particular case of interest is
that of the five-dimensional spacetime, where the moving domain wall may be interpreted as our braneworld universe. Thus in this subsection, we will focus on the five-dimensional case. The cosmological equations on the domain wall have been studied in \cite{gasa,kraus,ida}.

The behavior after the nucleation crucially depends on the sign of $A$ and the branch of the domain wall. For $A>0$, there are two branches of $r_{\ast,+}$ and $r_{\ast,-}$ (see Sec. 3.3).
They correspond to the local maximum and minimum of the potential $V(r)$, respectively. However, as we have discussed previously,
for the $(-)$-branch domain wall the BH pair cannot be formed
for $A>\frac{\Lambda}{6}$ obtained from Eq. (\ref{Acos})
with $n=4$. Thus, we focus on the $(+)$-branch domain wall.
In the (+)-branch, the domain wall is initially the Einstein static universe just after the nucleation, but may evolve once the wall position deviates from the equilibrium point due to the small perturbations. Assuming that the domain wall evolves as $r=a(\tau)$,
where $a(\tau)$ plays the role of the scale factor of the
Robertson-Walker universe on the domain wall and $\tau$ is now the proper time in the Lorentzian signature, the cosmological equation is given by $\dot{a}^2+V(a)=0$, hence
\bea
\frac{\dot{a}^2}{a^2}
+\frac{1}{a^2}
=\frac{1}{3}\lambda +\frac{2m_{\ast}}{a^4}
-\frac{q^2}{a^6},
\eea
where $\lambda:=3A$ becomes the effective cosmological constant \cite{gasa,kraus}. The second and third terms on the right-hand side
are induced due to the mass and charge of the created BHs, respectively. The mass term behaves as the radiation in the universe \cite{gasa,kraus}, while the charge term behaves as the stiff matter
with a negative energy density. Cosmological solutions of the domain wall universe in the charged BH background have been studied in e.g., Ref. \cite{mp01}. There are two possibilities, namely the expanding ($a>r_{\ast,+}$) or contracting ($a<r_{\ast,+}$) domain wall.
The first case corresponds to the expanding universe, approaching the de Sitter inflation with $\lambda$, since the contributions of the mass and charge terms are diluted. The second case corresponds to the initially collapsing universe. However, in this case the domain wall does not run into the singularity and experiences a bounce,
since there is the barrier in $V(a)$ because of the charge $q$.

For $A<0$, the BH pair cannot be formed for $|A|<\frac{|\Lambda|}{6}$ obtained from Eq. (\ref{Acar}) with $n=4$, and we do not consider this case.

\section{Summary and discussions}

In this paper, we have studied the creation of a BH pair separated by a domain wall in the four- or five-dimensional spacetimes with a cosmological constant. We have constructed the solution representing a BH pair with a domain wall. In any dimensions, the solution which involves the domain wall outside the outer BH horizon can be formed only at the local maximum of the potential, which has a smooth limit to the neutral case. The singularity of the domain wall universe with charged BHs, as distinct from that with neutral BHs, can be avoided.

There are two kinds of approach to the creation of a BH pair. One is the bounce approach. The other is the quantum cosmological approach. In the semiclassical approximation, the formula of the production rate without a prefactor in the bounce approach is equivalent to the formula in the quantum cosmological approach \cite{bcha}.

We have computed the Euclidean action for the above system in four- and five- dimensional spacetimes by employing the quantum cosmological approach. We then have compared our results of the charged case with those of the neutral case with a same cosmological constant. We find that the production rate of a charged BH pair is always smaller than that in the case of a neutral BH pair in both dimensions. We also have confirmed that the Euclidean action is always equal to the minus entropy of the created BHs, following the area law. We have given the first explicit proofs on the relation between the pair creation rate and the area of horizons of the created BHs in the four- and five-dimensional spacetimes.
The spacetime of the domain wall can be provided either by a cut-and-paste method or by the instanton solution mediating tunneling between the degenerate vacua in curved space \cite{hw, Lee:2008hz, lllo, lllo2}. The braneworldlike object can be obtained as the interpolating instanton solution by applying the mechanism in \cite{lllo, lllo2}. From this point of view, our present work can be the basic framework to make the spacetime, where we can describe the dynamics of a domain wall universe in the charged BH spacetime not only in the  Einstein-Maxwell theory in Refs.~\cite{mp01, km01} but also in the more general $U(1)$ gauge theories in Refs.~\cite{llm01, mh00, dm01}. With this motivation, the application to the braneworld cosmology has also been discussed. The four-dimensional geometry just after the nucleation is the Einstein static universe. The branch which is smoothly connected to the neutral case corresponds to the local maximum of the potential and eventually evolves, while the others are the local minimum.

In this paper, we have employed the time periodicity by the Hawking temperature of BHs \cite{hh00} so that no conical singularity appears at the horizon. The Euclidean geometry is smooth at the horizon, and therefore we only employed the boundary term by the presence of the domain wall. We omitted the boundary term for the form field, because it vanished in this paper.

We briefly discuss the possible generalization and the results expected from the previous results.

The first straightforward generalization is to the case of more than six-dimensional spacetimes. Even in such a case, the phase diagrams of the BHs and domain walls are very similar to Figs.\ 1-2 and Fig.\ 3, respectively. Similarly, the instanton for a pair of BH with a
domain wall can be constructed. Note that the horizon inside the domain wall can be formed only at the local maximum of the potential $V(r)$ for $A$ satisfying Eq. (\ref{Acos}). Then, the creation probability is also similarly evaluated, which gives the essentially the same results as in the previous cases. In particular, the Euclidean action for the charged BH pair with a domain wall is  expected to be
\bea
S_E(cbls)= -\frac{\cal A}{4G},
\eea
where ${\cal A}= 2\times \Omega_{n-1}r_+^{n-1}$ is the area of the outer horizon, where the factor 2 means two sides. Following the area law, $(-S_E)$ coincides with the BH entropy, which is expected from our construction of the instanton. The nucleation rate $e^{-S_E}$ for a charged BH pair is always suppressed in comparison with that of a neutral pair for a given cosmological constant, since the area of the outer horizon decreases as the charge increases.

Although it may be most plausible that two identical BHs are produced across the wall, it would be interesting to discuss the possibility without $Z_2$-symmetry. This is the case where the cosmological constant in both sides is different and the domain wall is interpolating two different vacua. In this case, in terms of the continuity of the magnetic field strength across the wall,
two BHs must have the same amount of charge. In the case with $Z_2$ symmetry, for a given cosmological constant $\Lambda$, charge $q$ and domain wall tension $\sigma$ (or $A$), we could determine the positions of the outer event horizon $r_+$ and that of the domain wall $r_{\ast}$, and the BH mass $m_{\ast}$, through the horizon condition $f(r_+)=0$ and two stationary conditions for the wall Eq.\ (\ref{wall_position}). We can extend this argument to the case without $Z_2$ symmetry. For the given bulk cosmological constant $\Lambda_I$, charge $q$, and tension of the domain wall $\sigma$,
where $I$ specifies the side with respect to the wall,
totally five quantities, i.e., the masses $m_I$ and horizon positions $r_{+,I}$ of BHs (four quantities) and the domain wall position $r_{\ast}$ must be specified by the five independent conditions.
In our case, they are the horizon condition in each side $f_I(r_{+,I})=0$ (hence two conditions), two stationary conditions as Eq.\ (\ref{wall_position}), and the continuity of $\beta \sqrt{f(r_{\ast})}$ across the wall. The last condition is due to the requirement that the temperature measured by an observer
on the domain wall must be unique. Thus, in contrast to the case with $Z_2$ symmetry, even if a solution for the above set of equations exists, the masses of BHs are generically different. Otherwise, the system becomes overdetermined. It would be interesting to look for instanton solutions satisfying the above five relations, and evaluate the nucleation probability, although we now leave these issues for future studies.

\section*{Acknowledgements}
We would like to thank Hongsu Kim for helpful discussions and comments. This work was supported by the Korea Science and Engineering Foundation (KOSEF) grant funded by the Korea government(MEST) through the Center for Quantum Spacetime(CQUeST) of Sogang University with grant number R11 - 2005 - 021.
WL was supported by the National Research Foundation of Korea Grant funded by the Korean Government (Ministry of Education, Science and Technology)[NRF-2010-355-C00017]. MM is grateful for the hospitality of the CQUeST.

\end{document}